\documentclass[conference]{IEEEtran}
\IEEEoverridecommandlockouts
\usepackage{amsmath}
\usepackage{color}\usepackage{bbm}
\usepackage{multirow}
\usepackage{amsmath}
\usepackage{mwe}
\usepackage{amsthm} 
\usepackage{amssymb}
\usepackage{float}
\usepackage{xurl}
\usepackage{nccmath}
\usepackage{graphicx}
\usepackage{subcaption}
\usepackage[linesnumbered,ruled]{algorithm2e}
\usepackage{booktabs}
\usepackage{mwe}
\usepackage{cite}
\usepackage{graphicx}
\usepackage{textcomp}
\usepackage[normalem]{ulem}
\usepackage{xcolor}
\usepackage{tabulary}
\usepackage{array} 
\usepackage{mathtools}

\usepackage{hyperref}
\definecolor{linkcolour}{rgb}{0,0.2,0.6}
\definecolor{xgreen}{rgb}{0.2,0.6,0.0}
\definecolor{xred}{rgb}{0.7,0.1,0.0}
\hypersetup{colorlinks,breaklinks,citecolor=red!50, urlcolor=linkcolour, linkcolor=xgreen}

\usepackage{amsmath,amssymb,amsfonts}
\usepackage{algpseudocode}
\usepackage{xspace}
\usepackage{kotex}
\newcommand{\BfPara}[1]{{\noindent\bf#1.}\xspace}

\def\BibTeX{{\rm B\kern-.05em{\sc i\kern-.025em b}\kern-.08em
    T\kern-.1667em\lower.7ex\hbox{E}\kern-.125emX}}
\begin{document}

\title{AoI-Aware Markov Decision Policies for Caching
}

\author{$^\dag$Soohyun Park, $^\S$Soyi Jung, $^\ddag$Minseok Choi, and $^\dag$Joongheon Kim
\vspace{2mm}
\\
$^\dag$\textit{Korea University}
\quad 
$^\S$\textit{Hallym University}
\quad
$^\ddag$\textit{Kyung Hee University}
}

\maketitle

\begin{abstract}
% To support rapid and accurate autonomous driving, road environment information, which is difficult to obtain through vehicle sensors itself, is collected and utilized through communication with surrounding infrastructure. We consider a scenario that utilizes infrastructure such as road side units (RSUs) with distributed cache and macro base station (MBS) which acts as an central database in situations where caching of road environment information is required.
% Due to the rapidly changed road environment, a concept which represents a freshness of the road content, age of information (AoI), is important.
% Based on the AoI value, it is essential to keep appropriate content in the RSUs in advance, update it before the content is expired ,and send the content to the vehicles which want to use it. 
% However, too frequent content transmission for the minimum AoI may lead to indiscriminate use of network resources. Therefore, it is important to find an appropriate compromise. 
% The object of this paper is reducing the system cost used for content delivery through the proposed system while minimizing the content AoI present in MBS, RSUs and UVs. We propose a new solution using Markov Decision Process (MDP) and Lyapunov control, respectively, which guarantee optimality. 
We consider a scenario that utilizes road side units (RSUs) as distributed caches in connected vehicular networks. %In the scenario, cache management to service connected vehicles is required. 
The goal of the use of caches in our scenario is for rapidly providing contents to connected vehicles under various traffic conditions.
During this operation, due to the rapidly changed road environment and user mobility, %a concept which represents a freshness of the road content, 
the concept of age-of-information (AoI) is considered for (1) updating the cached information as well as (2) maintaining the freshness of cached information. %Too frequent content transmission for the minimum AoI may lead to indiscriminate use of network resources. Therefore, it is significant to find an appropriate compromise. 
The frequent updates of cached information maintain the freshness of the information at the expense of network resources. Here, the frequent updates increase the number of data transmissions between RSUs and MBS; and thus, it increases system costs, consequently. Therefore, the tradeoff exists between the AoI of cached information and the system costs. 
Based on this observation, the proposed algorithm in this paper aims at the system cost reduction which is fundamentally required for content delivery while minimizing the content AoI, based on Markov Decision Process (MDP) and Lyapunov optimization. 

\end{abstract}

%\begin{IEEEkeywords}
%Caching system, Age-of-information, Connected vehicle, Distributed edge, MDP, Lyapunov optimization
%\end{IEEEkeywords}

\section{Introduction}
\BfPara{Motivation} Smart vehicles intelligently interact with their surroundings in real-time to determine the optimal driving decisions and assist drivers for safe and fast driving. For the purpose of rapid and accurate driving decisions to ensure the driving stability in fast-moving vehicle network environments, many research contributions are proposed including data transmission control considering the data freshness, e.g., age-of-information (AoI)~\cite{aoin}. 
%The transmitted data freshness affects to driving control decision of the smart vehicle. 
Because road-side traffic information is delay-sensitive for connected vehicles, the data freshness is critical for optimizing the performance of link connectivity in vehicular networks. 
Therefore, cache-assisted connected vehicles have been widely considered for delay-sensitive data delivery and AoI-aware fresh content management. %and sharing based on the concept of AoI using the infrastructure of the vehicle network is especially important.

\BfPara{Related Work}
The AoI is a metric for information freshness that measures the time that elapses since the last received fresh update was generated at the source~\cite{TIT[11]}. 
In the environment where data updates are required (e.g., mobile device's recent position, speed, and other control information), the analysis and optimization of AoI have been extensively studied in various scenarios~\cite{TIT_AoI}. 
Moreover, the applications and usages of the concept of AoI are also widely studied, e.g., ultra-reliable vehicular communications~\cite{TCOM[15]}.
Lastly, AoI is utilized as an important performance evaluation criteria in random access and caching replacement research~\cite{TCOM_AoI_caching}.

\BfPara{Contribution}
We propose a joint cache replacement and content delivery scheme in cache-assisted connected vehicles that minimizes the network cost while limiting the AoI of traffic conditions.
The proposed scheme adaptively controls the tradeoff between the content AoI and network resource consumption, depending on rapidly changing road environments, user mobility, as well as the AoI of contents.
%We optimize content transmission decisions for the distributed cache management considering data freshness in the system. In the proposed connected vehicle network, data delivery reflecting rapidly changing road environment information is essential, so caching research considering AoI is important. We propose a connected vehicle system that considers not only the content AoI but also waiting times and communication cost. 

\section{AoI-aware MDP for Caching}
In our reference network model, we assume cache replacements of roadside units (RSUs) and content delivery from RSUs to user vehicles (UVs) can be separately performed. 
%we consider that content transmission which contains content update for roadside units (RSUs), and content service to user vehicles (UVs) is achieved in one time slot independently. We present the optimal solution for each transmission stage.
\subsection{Overall Architecture}

There are $N_u$ UVs, $N_R$ RSUs, and one macro base station (MBS) are deployed around the straight road. The road has $L$ regions, the conditions of the regions are not all the same and one content is created for each region. %The UVs move in one direction, and the state of the road through which each vehicle (UV or CV) passes is different for each region.
MBS is in the center of the entire road, and RSUs which cover $L'$ regions are distributed at specific distance intervals.
%The CVs generate content for each region when they pass through the region and deliver specific content when connected to MBS. 
The UVs, which are thead-hoc smart connected vehicles, move in one direction and request the RSU for the contents what they need and receive necessary information through them.
Each RSU is a service provider that delivers content directly to the UVs.
Because contents cached in the RSU get older, cache management is necessary to maintain the freshness of cached contents. %so that RSU has the latest data.
Suppose that the MBS has all the new contents generated at each time slot, and observes the cache states of all RSUs; therefore, it periodically decides what contents and which RSUs have to be updated for providing the fresh information to UVs. %before the AoI of the content in the RSU cache is old.
In addition, we assume that all contents have the same file size and different maximum AoI value limits.

\subsection{Cache Management via Markov Decision Process (MDP)}
For optimizing cache management of RSUs by the MBS, maximizing a total utility which is the summation of AoI value and communication cost. We define our own MDP model with the definitions of states $\mathcal{S}$, actions $\mathcal{A}$, and $\mathcal{R}$, as follows. %guarantees the optimal solution for every moment. 
%We characterize the content caching network environment as follows:
\begin{itemize}
\item \BfPara{State ($\mathcal{S}$):} 
%Information used by MBS in an environment to which MDP is applied, is described. 
The state contains AoI of all contents in the system and the content population that each RSU has. AoI information includes maximum AoI value and AoI of the contents which are stored in MBS and RSUs.
% \begin{equation}
%     \mathcal{S}(t)=\{[A(t)], [d(t)], [h(t)], [p(t)]\}
% \end{equation} 
% $A^c_{j,h}(t),A_h(t),A^R_{k,h}(t),A^{max}$: AoI value for content $h$ stored in CV$j$, MBS and RSU$k$ depending on the action $x$ and $y$. $A^{max}$ is maximum AoI value applied equally to CV$j$, MBS and RSU$k$.\\
% $d_j(t), d_k(t)$: Distance from MBS to CV$j$ and RSU$k$.\\
% $h(t)$: Channel state of MBS which is determined by action $x$ and $y$.\\
%  $p^k_h(t)$: The popularity of RSU$k$'s content$h$.\\
\item \BfPara{Action ($\mathcal{A}$)} In MDP, there is only one action variable. That means whether the content in the RSU is updated or not by MBS. The action is expressed as binary, when the variable has $1$, the content update is occurred. Each RSU has several contents and only one content is updated at a time. The action is denoted as $x^k_h(t)$, $k$ and $h$ indices are for RSUs and contents in an RSU.
% \begin{equation}
%     \mathcal{A}(t) = \{[x(t)], [y(t)]\}
% \end{equation}
% $x^j_h(t)$: Whether the content $h$ is uploaded or not from CV$_j$ to MBS\\
% $y^k_h(t)$: Whether the content $h$ in RSU$k$ is updated or not by MBS \\
\item \BfPara{Reward ($\mathcal{R}$)} The reward function is the combination of AoI utility of RSUs weighted by $w$ and the network cost of the MBS for cache replacements of RSUs (negative reward), as, % described follows: %which are determined by the action variable.

\begin{eqnarray}
    \mathcal{U}(t) &=& \mathcal{U}^{RSU}_{AoI}(t)\cdot w - \mathcal{U}^{MBS}_{cost}(t) \label{eq:utility}\\
    \mathcal{U}^{RSU}_{AoI}(t) &=& \sum^{N_R}_{k=1}\nolimits\sum^{L}_{h=1}\nolimits \frac{A^{\max}_h}{A^R_{k,h}(x^k_h(t))}p^k_h(t)  \label{eq:stage1 AoI utility}\\
    \mathcal{U}^{MBS}_{cost}(t) &=& \sum^{N_R}_{k=1}\nolimits\sum^{L}_{h=1}\nolimits C^k_h(x^k_h(t))  \label{eq:stage1 cost utility}
\end{eqnarray}
where 
$\mathcal{U}^{RSU}_{AoI}(t)$ means the proportion of the current AoI value of the RSU $A^R_{k,h}(t)$ to the reference maximum AoI $A^{\max}_{h}$. When the action variable is $1$, the old content of RSU is replaced by the new version which is stored in the MBS.
% \begin{multline}
%     A^R_{k,h}(t) = (1-y^k_h(t))\cdot(A^R_{k,h} (t-1) + 1) \\+ y^k_h(t)\cdot A_h(t-1), \forall k, h\in N_R, L
%     \label{eq:AoI_RSU}
% \end{multline}
% \begin{multline}
%     A_h(t) = \sum_{\forall j\in N_C}\{(1-x^j_h(t))\cdot(A_h(t-1)+1) \\+ x^j_h(t)\cdot A^c_{j,h}(t-1)\}, \forall h\in L
%     \label{eq:AoI_MBS}
% \end{multline}
$\mathcal{U}^{MBS}_{cost}(t)$ is the negative utility which represents the communication resource using cost $C^k_h(t)$ for the $h$-th content in $k$-th RSU. 
\end{itemize}
%We apply the content popularity at time step $t$. Its intention is to ensure that even if frequent communication occurs for the freshness of content, if the content is popular from UVs that the RSU should service, the negative utility for communication cost has a smaller value than other cases.

\subsection{Lyapunov-based Service Control}
For the UV service optimization of each RSU, the object is meeting trade-off between latency of UV ($Q[t]$) which can be represented as queue and RSU communication cost (i.e., $C(\alpha[t])$) before UV leaves the RSU coverage. At the same time, guaranteeing the valid content service before the waiting time is expired is also considered. We propose an optimization algorithm based on Lyapunov optimization~\cite{tmc201907koo}, as follows,
\begin{equation}
    \min: \lim_{T\rightarrow\infty}\frac{1}{T}\sum_{t=1}^{T}\nolimits C(\alpha[t])\label{eq:opt}
\end{equation}
subject to queue stability, i.e., $\lim_{T\rightarrow\infty}\frac{1}{T}\sum_{t=1}^{T}\nolimits Q[t]<\infty$ and AoI requirements, i.e., $\sum_{h=1}^{L'}\nolimits A(\alpha[t])\leq A^{\max}_h$.
Based on Lyapunov control~\cite{jsac201806choi}, the optimal decision whether to service the UV at this time is as~\eqref{eq:optimal_action},
\begin{equation}
    \alpha^{*}[t] \leftarrow \arg\min_{\alpha[t]\in\mathcal{S}} \left[V\cdot C(\alpha[t]) - Q[t]b(\alpha[t])\right]
    \label{eq:optimal_action}
\end{equation}
where $\alpha^{*}[t]$, $\mathcal{S}$, $V$, and $b(\alpha[t])$ are optimal decision at $t$, set of all possible decisions $\alpha[t]$, tradeoff co-efficient, and departure (i.e., processing speed) with $\alpha[t]$, respectively.

In order to verify whether \eqref{eq:optimal_action} works as desired, it can be confirmed by evaluating following two extreme two cases.
%Then, the \eqref{eq:lyapunov-final} tries to minimize $V\cdot C_i(\alpha_i[t])$, i.e., the RSU dose not allocate channel to UV$_i$ in a situation where the algorithm satisfies with condition ~\eqref{eq:Lyapunov_AoI_const} due to the waiting time of the UV is not so long there is enough time to wait. This is semantically true because we can focus on the main objective, i.e., communication cost of the RSU, because stability is already achieved at this moment.
If $Q[t] = 0$, the~\eqref{eq:optimal_action} tries to minimize communication cost, i.e., the RSU dose not service the UV. This is semantically true because we can focus on the main objective, because stability is already achieved.  
On the other hand, if $Q[t] \approx \infty$, the~\eqref{eq:optimal_action} will focus on maximizing $b(\alpha[t])$ by deciding UV service at that time. The queue which has the accumulated time is emptied by $b(\alpha[t])$ and the queue pursues stability.

\section{Performance Evaluation}
% \begin{figure}[t]
%     \begin{center}
%         \includegraphics[width=0.75\linewidth]{lyapunov.png}
%     \end{center}
%     \caption{Simulation results}
%     \label{fig:results}
% \end{figure}

\begin{figure}%
    \subfloat[AoI-aware content caching.]{{\includegraphics[height=0.15\textwidth ]{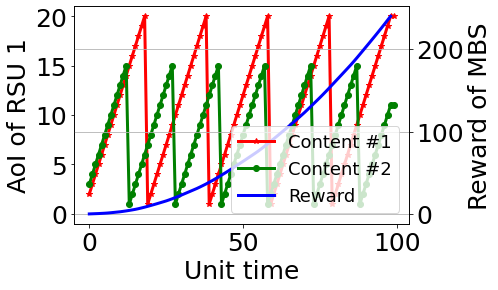} }\label{fig:MDP}}%
    \subfloat[Delay-aware content service.]{{\includegraphics[height=0.15\textwidth ]{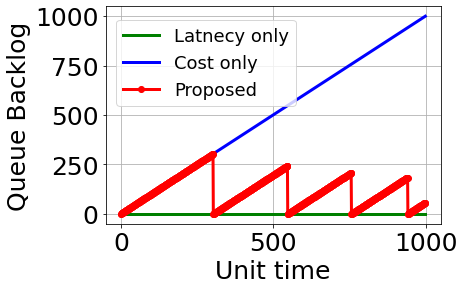} }\label{fig:Lyapunov}}
    \caption{Simulation results.}
    \label{fig:results}
    \vspace{-5mm}
\end{figure}

The road environment is divided by several regions and has a different state, and each state has a different maximum value. All regions are covered by $5$ RSUs, and only the content of the region covered by the RSU is cached. In this simulation, the initial content AoI value of the MBS and RSU, and the status for each region are determined as random. Similarly, the content requested by the UV to the RSU is randomly generated.
Performance evaluations based on MDP and Lyapunov optimization are conducted by $1000$ iteration, respectively.
Fig. \ref{fig:MDP} shows the content AoI change and the cumulative reward. Here, there are $4$ RSUs and each RSU has $5$ cached content. Totally, $20$ contents are managed by MBS, each content is updated before the AoI value exceeds the maximum $A^{\max}_h$. Instead of showing the AoI changes of all contents, we select two contents in the cache of RSU $1$ and show them over time unit. In Fig.~\ref{fig:MDP}, we check the cumulative reward of MBS by the proposed update decision also continues to rise. The reward is obtained by managing $4$ RSUs' cache and calculated by ~\eqref{eq:utility} based on the AoI utility of RSUs and the network cost of the MBS.
Fig. \ref{fig:Lyapunov} shows the latency of UV ($Q[t]$). There is a service decision in the RSU at an appropriate time to satisfy the stability of the queue.
That means the proposed method considers trade-off between cost and latency compared to the other two algorithms.

\section{Concluding Remarks}
This paper proposed a two-stage joint AoI-aware cache management and content delivery scheme for providing fresh road contents to connected vehicles. 
We present an MDP-based algorithm for cache management of RSUs to limit the AoI of cached contents.
In addition, the content delivery from cache-enabled RSUs to UVs is adaptively optimized depending on the current AoI of contents and rapidly time-varying traffic contisions under the Lyapunov-based control.
%Optimization of transmission decisions at each stage for distributed cache management and content service using road infrastructure are essential. In this paper, therefore, a new dynamic decision algorithms are proposed and they are based on Markov Decision Process (MDP) and Lyapunov optimization.

\BfPara{Acknowledgment}
This research was supported by NRF-Korea \& ITRC (2021R1A4A1030775 \& IITP-2022-2017-0-01637). S. Jung, M. Choi, and J. Kim are corresponding authors (e-mails: sjung@hallym.ac.kr, choims@khu.ac.kr, joongheon@korea.ac.kr).


\bibliographystyle{IEEEtran}

\vspace{12pt}

\end{document}